\documentclass[twocolumn,english]{revtex4-1}
\usepackage[T1]{fontenc}
\usepackage[latin9]{inputenc}
\usepackage{amssymb}
\usepackage{graphicx}

\makeatletter
\usepackage{hyperref}

\makeatother

\usepackage{babel}
\begin{document}
\title{Learning phase transitions from regression uncertainty: A new regression-based
machine learning approach for automated detection of phases of matter}
\author{Wei-Chen Guo}
\email{weichen.guo@m.scnu.edu.cn}

\affiliation{Institute for Theoretical Physics, School of Physics, South China
Normal University, Guangzhou 510006, China}
\affiliation{Guangdong Provincial Key Laboratory of Quantum Engineering and Quantum
Materials, Guangdong-Hong Kong Joint Laboratory of Quantum Matter,
South China Normal University, Guangzhou 510006, China}
\author{Liang He}
\email{liang.he@scnu.edu.cn}

\affiliation{Institute for Theoretical Physics, School of Physics, South China
Normal University, Guangzhou 510006, China}
\affiliation{Guangdong Provincial Key Laboratory of Quantum Engineering and Quantum
Materials, Guangdong-Hong Kong Joint Laboratory of Quantum Matter,
South China Normal University, Guangzhou 510006, China}
\begin{abstract}
For performing regression tasks involved in various physics problems,
enhancing the precision or equivalently reducing the uncertainty of
regression results is undoubtedly one of the central goals. Here,
somewhat surprisingly, we find that the unfavorable regression uncertainty
in performing the regression tasks of inverse statistical problems
actually contains hidden information concerning the phase transitions
of the system under consideration. By utilizing this hidden information,
we develop a new unsupervised machine learning approach for automated
detection of phases of matter, dubbed learning from regression uncertainty.
This is achieved by revealing an intrinsic connection between regression
uncertainty and response properties of the system, thus making the
outputs of this machine learning approach directly interpretable via
conventional notions of physics. We demonstrate the approach by identifying
the critical points of the ferromagnetic Ising model and the three-state
clock model, and revealing the existence of the intermediate phase
in the six-state and seven-state clock models. Comparing to the widely-used
classification-based approaches developed so far, although successful,
their recognized classes of patterns are essentially abstract, which
hinders their straightforward relation to conventional notions of
physics. These challenges persist even when one employs the state-of-the-art
deep neural networks that excel at classification tasks. In contrast,
with the core working horse being a neural network performing regression
tasks, our new approach is not only practically more efficient, but
also paves the way towards intriguing possibilities for unveiling
new physics via machine learning in a physically interpretable manner.
\end{abstract}
\maketitle

\section{Introduction}

In recent years, the artificial neural network (NN) based machine
learning techniques have stimulated the rapid development of automated
data-driven approaches to investigate a wide range of physics problems
\citep{Carleo_RMP_2019,Carrasquilla_AdvPhysX_2020,Volpe_Nat_Mach_Intell_2020}.
As one of the central classes of problems in condensed matter physics
and statistical physics, classifying phases of matter and identifying
phases transitions is a major focus of applying both generative machine
learning \citep{Wetzel_PRE_2017,Kai_Zhou_CPL_2022,Kai_Zhou_arXiv_2020,DAngelo_PRR_2020,Singh_SciPostPhys_2021}
and discriminative machine learning \citep{Melko_Nat_Phys_2017,van_Nieuwenburg_Nat_Phys_2017,Venderley_PRL_2018,Guo_PRE_2021,Melko_PRB_2018,Suchsland_PRB_2018,Lee_PRE_2019,Chernodub_PRD_2020,Schindler_PRB_2017,Das_Sarma_PRL_2018,Khatami_PRX_2017,Broecker_Sci_Rep_2017}.
Particularly, approaches utilizing the power of NNs in performing
classification discriminative tasks have been developed and succeeded
in providing data-driven evidence on the existence of various phase
transitions \citep{Melko_Nat_Phys_2017,van_Nieuwenburg_Nat_Phys_2017},
including the phase transitions associated with nonequilibrium self-propelled
particles \citep{Venderley_PRL_2018,Guo_PRE_2021}, topological defects
\citep{Melko_PRB_2018,Suchsland_PRB_2018,Lee_PRE_2019,Chernodub_PRD_2020},
many-body localization \citep{Schindler_PRB_2017,Das_Sarma_PRL_2018},
strongly correlated fermions \citep{Khatami_PRX_2017,Broecker_Sci_Rep_2017},
etc. And besides the widely-involved classification tasks, there is
yet another fundamental class of discriminative tasks that can be
dealt with efficiently by machine learning, namely, the regression
tasks \citep{Goodfellow_Book_2016}. In fact, applying the machine
learning techniques designed for regression tasks to physics problems
is now giving rise to a burgeoning field towards automated theory
building, where, for instance, the symbolic regression \citep{Tegmark_SciAdv_2020,Tegmark_PRE_2019,Tegmark_PRE_2021b}
has been successfully applied to extract the equations of motion \citep{Tegmark_PRE_2021a},
symmetries \citep{Tegmark_PRL_2022}, and conservation laws \citep{Tegmark_PRL_2021}
from various types of data of physical systems.

But despite the exciting development in this context, a quite natural
application scenario of machine learning techniques has received little
attention so far, namely, classifying phases of matter and identifying
phases transitions by utilizing the power of NNs in performing regression
tasks \citep{Schafer_PRE_2019,Greplova_NJP_2020,Arnold_PRR_2021,Arnold_PRX_2022}.
Currently, to classify phases and identify phase transitions with
machine learning, the widely-used approaches usually train the NN
to perform a certain classification task \citep{Melko_Nat_Phys_2017,van_Nieuwenburg_Nat_Phys_2017}.
In spite of their successes, they could generally face the lack of
interpretability as the classes of patterns recognized by them are
essentially abstract, hence cannot assume straightforward relation
to conventional notions of physics \citep{Gokmen_PRL_2021_RSMI,Gokmen_PRE_2021_RSMI,Lei_Wang_PRB_2019,Lei_Wang_PRL_2018,YiZhuang_You_PRR_2020,Kim_Nat_Commun_2021,Kim_arXiv_2021}.
For instance, the supervised learning-with-blanking approach \citep{Melko_Nat_Phys_2017,van_Nieuwenburg_Nat_Phys_2017}
investigates phase transitions by examining the intersection of the
NN's binary classification confidences. But to relate this intersection,
which separates two recognized classes, to the actual phase transition
point between two distinct phases of matter, it requires additional
system-specific knowledge, such as the number of existing phases,
the absence or presence of phase separation and crossover, etc. Similarly,
the unsupervised learning-by-confusion approach \citep{van_Nieuwenburg_Nat_Phys_2017}
and the unsupervised scanning-probe approach \citep{Guo_EPL_2021}
investigate phase transitions by contrasting the NN's recognition
performance in the binary classification tasks with different assumed
targets. Yet, relating the optimal target to a specific phase transition
still necessitates external physical information beyond the scope
of these approaches. The challenges persist even when one employs
the state-of-the-art deep NNs that excel at classification tasks.
While for the NNs performing regression tasks, in contrast, the readily
interpretable meaning of their outputs can usually be traced back
to the regression tasks themselves straightforwardly. Taking the NNs
designed for solving the inverse statistical problem (ISP) \citep{Nguyen_Adv_Phys_2017}
for example, they are trained to perform regression tasks of reconstructing
certain system parameters from given samples of observed system configurations,
with their outputs naturally holding the same meaning as the system
parameters they try to reconstruct. In these regards, the power of
NNs in performing regression tasks might shed new light on automated
detection of phases of matter. Introducing a novel regression-based
approach offers opportunities to complement the widely-used classification-based
approaches by enhancing the interpretability of results from a physical
standpoint. This thus raises the fundamental and intriguing question
of whether and how phase transitions can be unsupervisedly revealed
by NN-based regression.

Here, this question is addressed for the prototypical regression tasks
in ISP performed by NNs. To date, significant efforts have been made
to apply machine learning techniques to ISP with a primary focus on
accurately and precisely inferring system parameters in various models,
such as inferring the coupling strengths in spins systems \citep{Nguyen_Adv_Phys_2017,Aurell_PRL_2012,Nguyen_PRL_2012,Decelle_PRE_2016,Donner_PRE_2017,Periwal_PRE_2020,Pan_Zhang_PRL_2019},
in monomer-dimer systems \citep{Contucci_J_Phys_A_Math_Theor_2017},
and in restricted Boltzmann machine \citep{Beentjes_PRE_2020}. Other
applications involved inferring the single-particle spectral density
function from Green's functions \citep{Fournier_PRL_2020}, and inferring
the disorder potential from quantum transport properties \citep{Percebois_PRB_2021},
etc. In these previous investigations, enhancing the precision or
equivalently reducing the uncertainty of the regression results was
regarded as one of the central goals. But somewhat surprisingly, we
find that this unfavorable regression uncertainty actually contains
hidden information that can be utilized to reveal possible phase transitions.
In this work, these findings shall be demonstrated in the ferromagnetic
Ising model and $q$-state clock models. The former exhibits a second-order
phase transition, whose critical temperature is $T_{C}=2\slash\ln(1+\sqrt{2})\,(J\slash k_{B})$
\citep{Kogut_RMP_1988}. The $q$-state clock models also exhibit
this transition with small $q$ (e.g., $q=3$), while with slightly
larger $q$ (e.g., $q=6,7$), it exhibits two Berezinskii-Kosterlitz-Thouless
(BKT) phase transitions with an intermediate phase \citep{Kosterlitz_Rep_Prog_Phys_2016,Kosterlitz_JPhysC_1973,Kosterlitz_JPhysC_1974}.
These many-body models of interacting spins are not only important
in condensed matter physics and statistical physics, but in recent
years also typically employed to examine the ability of new machine
learning approaches \citep{Carleo_RMP_2019,Carrasquilla_AdvPhysX_2020,Volpe_Nat_Mach_Intell_2020}.

Taking the ISP associated with the Ising model (also known as the
inverse Ising problem \citep{Nguyen_Adv_Phys_2017,Aurell_PRL_2012,Nguyen_PRL_2012,Decelle_PRE_2016,Donner_PRE_2017,Periwal_PRE_2020,Pan_Zhang_PRL_2019})
for instance, it asks the question of what the possible temperature
is for a given spin configuration. For any spin configuration generated
probabilistically, it does not correspond to an unique temperature,
which thus gives rise to intrinsic regression uncertainty for the
reconstructed temperatures {[}see Eq.~(\ref{eq:Output_variance})
and the error bars in the inset of Fig.~\ref{fig:Ising}(b){]}. As
demonstrated in this work in the cases of the ferromagnetic Ising
model (see~Fig.~\ref{fig:Ising}) and the three-state clock model
{[}see~Fig.~\ref{fig:Clock}(b){]}, the position of the non-trivial
minimum of regression uncertainty directly corresponds to the critical
point of the continuous phase transition of the system. It is further
shown that the regression uncertainty for reconstructing temperatures
of the six-state and seven-state clock models {[}see~Figs.~\ref{fig:Clock}(c)
and \ref{fig:Clock}(d){]} assumes two non-trivial minima, respectively,
which can provide a new type of data-driven evidence on the existence
of the intermediate phase in these systems. These findings clearly
suggest that the regression uncertainty in ISP generally contains
non-trivial hidden information. By utilizing this hidden information,
in this work we develop a new unsupervised machine learning approach
for automated detection of phases of matter, dubbed learning from
regression uncertainty (LFRU), which is distinguished from the various
classification-based machine learning approaches developed so far
\citep{Melko_Nat_Phys_2017,van_Nieuwenburg_Nat_Phys_2017,Venderley_PRL_2018,Guo_PRE_2021,Melko_PRB_2018,Suchsland_PRB_2018,Lee_PRE_2019,Chernodub_PRD_2020,Schindler_PRB_2017,Das_Sarma_PRL_2018,Khatami_PRX_2017,Broecker_Sci_Rep_2017,Xuefeng_Zhang_PRB_2019,Rui_Zhang_PRB_2019,Canabarro_PRB_2019,Kahng_PRR_2021,Guo_EPL_2021}.
This is achieved by revealing an intrinsic connection between regression
uncertainty and response properties of the system, thus making the
outputs of this machine learning approach directly interpretable via
conventional notions of physics (noticing however that the employed
NNs themselves still remain not interpretable). Moreover, since the
implementation of this approach is robust against different choices
of the NN architecture, it is expected that state-of-the-art developments
in the field of artificial intelligence and data science can be readily
exploited to automatically detect phases of matter in a physically
interpretable manner via LFRU.

\section{Identify phase transitions from regression uncertainty\label{sec:2phases}}

To see concretely how the regression uncertainty in ISP can be utilized
to reveal possible phase transitions, let us start with the ISP in
the prototypical ferromagnetic Ising model $H=-J\sum_{\langle i,j\rangle}s_{i}s_{j}$,
where the Ising spins $s_{i}=\pm1$ are located on a square lattice
with linear size $L$ and periodic boundary condition imposed. The
coupling strength $J=1$ is set as the energy unit in the following.

As the inverse of generating spin configurations that satisfy the
probability distribution $\exp(-H\slash T)\slash Z$ ($Z\equiv\sum_{\{s_{i}\}}\exp(-H\slash T)$
and the Boltzmann constant $k_{B}$ is set to be $1$) at a given
temperature $T$, the ISP in this case asks the question of what the
possible temperature is for a given spin configuration {[}see~Fig.~\ref{fig:Ising}(a)
for instance{]}, hence naturally corresponds to a prototypical regression
task. This regression task is performed by training the NN as an $L\times L\rightarrow1$
map from the set of spin configurations to the set of reconstructed
temperatures denoted by $T_{R}$. During the training process of the
NN, a large number of its parameters are optimized by minimizing a
loss function. The mean square error (MSE) \citep{Goodfellow_Book_2016}
is employed as the loss function $\mathbb{L}$ here for instance,
i.e.,
\begin{equation}
\mathbb{L}=\langle(T-T_{R})^{2}\rangle,\label{eq:MSE}
\end{equation}
where $\langle\cdot\rangle$ denotes the average over all the samples
in the training dataset (see Appendix \ref{sec:Training} and \ref{sec:Loss-and-maxima}
for more technical details concerning the data generation, NN training,
the architecture of NN, and the loss function).

\subsection{Hidden information in regression uncertainty}

\begin{figure}
\noindent \begin{centering}
\includegraphics[width=3.3in]{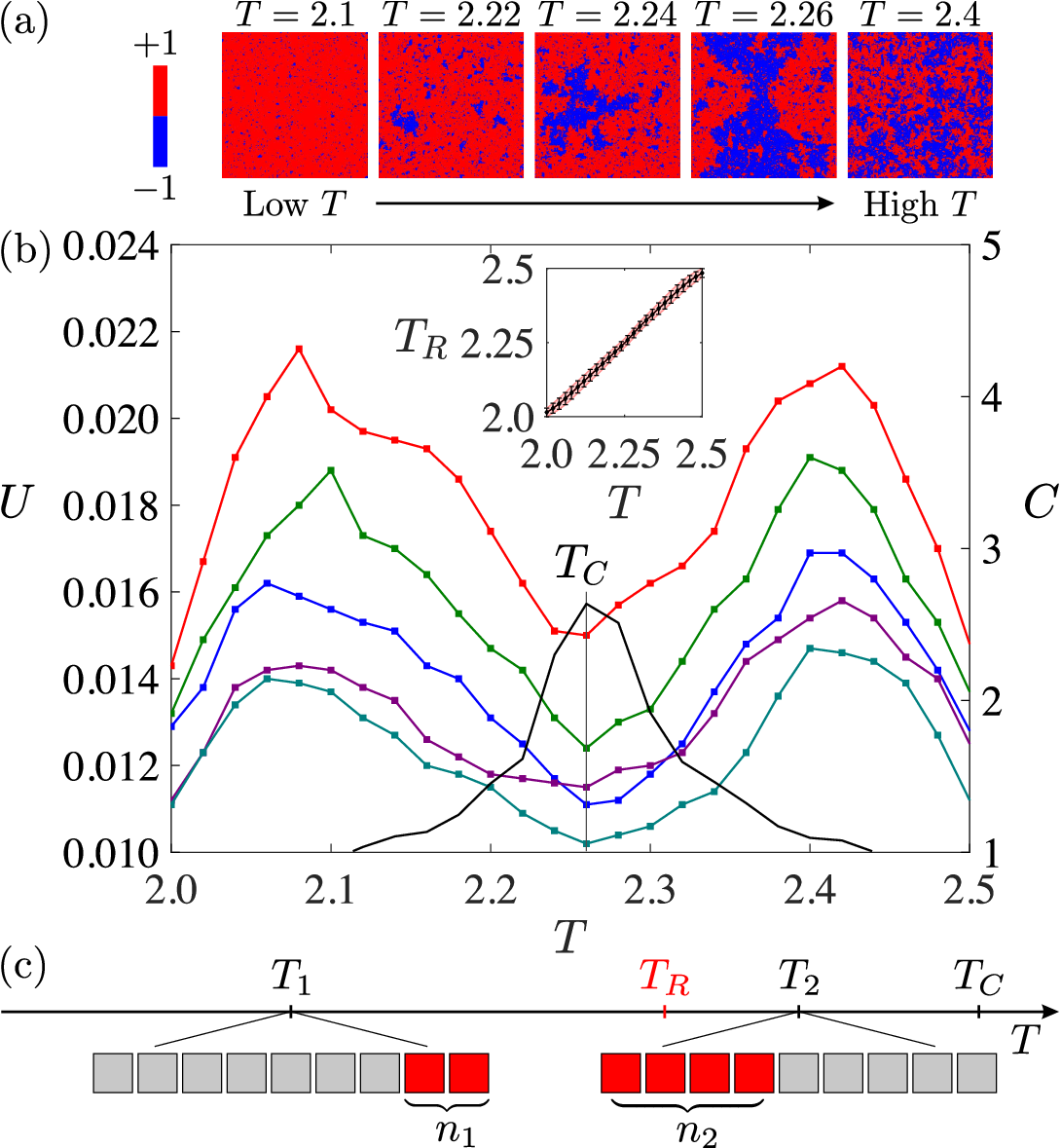}
\par\end{centering}
\caption{\label{fig:Ising}Learning the Ising transition from regression uncertainty.
(a) Typical real-space configurations (samples processed directly
by the NN) of the ferromagnetic Ising model. (b) Temperature dependence
of regression uncertainty $U(T)$ for reconstructing temperatures
$T\in[2.0,2.5]$ ($\Delta T=2\times10^{-2}$) of Ising model with
$L=120$ (red curve), $L=140$ (green curve), $L=160$ (blue curve),
$L=180$ (purple curve), and $L=200$ (cyan curve). The curve of regression
uncertainty $U(T)$ assumes a non-trivial M-shape and the valley appears
at $T=2.26\pm0.01$ which matches the critical temperature $T_{C}=2\slash\ln(1+\sqrt{2})\approx2.269$,
suggesting that it contains non-trivial information, i.e., the position
of the valley, which can be utilized to reveal possible phase transitions.
The black curve corresponds to the temperature dependence of the system's
heat capacity $C(T)$ with $L=180$. Inset: Regression results of
a well-trained NN for the corresponding ISP with $L=120$. The reconstructed
temperature $T_{R}$ is very close to the target $T$ (the diagonal
line represents the ideal regression results $T_{R}=T$ for the ISP),
but there always exists an intrinsic regression uncertainty $U(T)$
as shown by the error bars and the range of the colored error band.
(c) Schematic illustration of the origin of the hidden information
in regression uncertainty. In a simple scenario of reconstructing
two temperatures $T_{1}$ and $T_{2}$ ($T_{1}<T_{2}<T_{C}$), which
involves two sets of an equal number of samples (illustrated by squares)
generated probabilistically at $T_{1}$ and $T_{2}$, respectively,
there can be $n_{1}$ samples in the set of $T_{1}$ with relatively
high energy that are essentially indistinguishable from $n_{2}$ samples
in the set of $T_{2}$ with relatively low energy. To deal with these
$n_{1}+n_{2}$ indistinguishable samples (illustrated by red squares),
a simple but still decent way for the NN to accomplish its regression
task is to associate them with their weighted average temperature.
As a result, their $T_{R}$ tends to be closer to $T_{2}$ rather
than $T_{1}$ due to $n_{2}>n_{1}$ since the heat capacity that characterizes
the strength of thermal fluctuations satisfies $C(T_{2})>C(T_{1})$
in this case, which leads to $U(T_{2})<U(T_{1})$. This indicates
that the temperature at which $C(T)$ reaches its maximum should match
exactly the temperature at which $U(T)$ reaches its minimum as shown
in (b). See text for more details.}
\end{figure}

Due to the intrinsic statistical characteristic of ISP, for any set
of spin configurations, i.e., samples, generated at a given temperature
$T$ according to the probability distribution $\exp(-H\slash T)\slash Z$,
some of its spin configurations can also appear in other sets of spin
configurations generated at temperatures that are different from $T$.
As a direct result, the reconstructed temperatures $T_{R}$ for each
spin configuration in the same set generally does not assume the same
value. This thus gives rise to an intrinsic uncertainty of the regression
results. Straightforwardly, one can use the standard deviation $U(T)$
of well-trained NN's outputs to characterize this regression uncertainty
at a given temperature $T$,
\begin{equation}
U(T)\equiv\sqrt{\langle(T_{R}-\langle T_{R}\rangle_{T})^{2}\rangle_{T}},\label{eq:Output_variance}
\end{equation}
where $\langle\cdot\rangle_{T}$ denotes the average over all the
samples generated at $T$ in the test dataset. For instance, as one
can see from the error bars and the range of the colored error band
in the inset of Fig.~\ref{fig:Ising}(b), although the regression
uncertainty $U(T)$ can be generally very small for the well-trained
NN that assumes good performance in reconstructing the temperature,
it always exists.

Naturally, the regression uncertainty is unfavorable for reconstructing
the precise temperature $T$ from given spin configurations. But somewhat
surprisingly, we find that this unfavorable regression uncertainty
actually contains hidden information that can be utilized to identify
the critical point $T_{C}$ of the ferromagnetic phase transition
of the system. As one can see from Fig.~\ref{fig:Ising}(b), the
temperature dependence of regression uncertainty $U(T)$ assumes a
non-trivial M-shape for reconstructing temperatures $T\in[2.0,2.5]$
with the temperature spacing $\Delta T=2\times10^{-2}$. In particular,
one can find that its valley appears at $T=2.26\pm0.01$ (the resolution
is determined by $\Delta T$, see Appendix \ref{sec:Finite_size_analysis}
for a finite-size analysis with a higher resolution), which matches
the critical temperature $T_{C}=2\slash\ln(1+\sqrt{2})\approx2.269$
\citep{Kogut_RMP_1988} of the two-dimensional ferromagnetic Ising
model very well. This strongly suggests that the temperature dependence
of regression uncertainty $U(T)$ contains non-trivial information,
i.e., the position of the valley, which can be utilized to reveal
possible phase transitions.

\subsection{Connection between regression uncertainty and response properties}

To gain more intuitions for understanding the behavior of regression
uncertainty shown in Fig.~\ref{fig:Ising}(b) and see the possible
origin of its non-trivial information concerning phase transitions,
let us consider the training process of the NN in a simple scenario
where it is fed with only two sets of an equal number of samples generated
at temperatures $T_{1}$ and $T_{2}$ near the critical temperature
$T_{C}$ with $T_{1}<T_{2}<T_{C}$. Since the samples {[}illustrated
by squares in Fig.~\ref{fig:Ising}(c){]} in these two sets are generated
probabilistically, one expects that certain amount of samples, say
$n_{1}$ samples, that correspond to relatively high energy in the
set of $T_{1}$ are essentially indistinguishable from certain amount
of samples, say $n_{2}$ samples, that correspond to relatively low
energy in the set of $T_{2}$ {[}illustrated by red squares in Fig.~\ref{fig:Ising}(c){]}.
These essentially indistinguishable samples make it impossible for
any algorithm to perform the ISP regression perfectly (corresponding
to $\mathbb{L}=0$).

\begin{figure}
\noindent \begin{centering}
\includegraphics[width=3.3in]{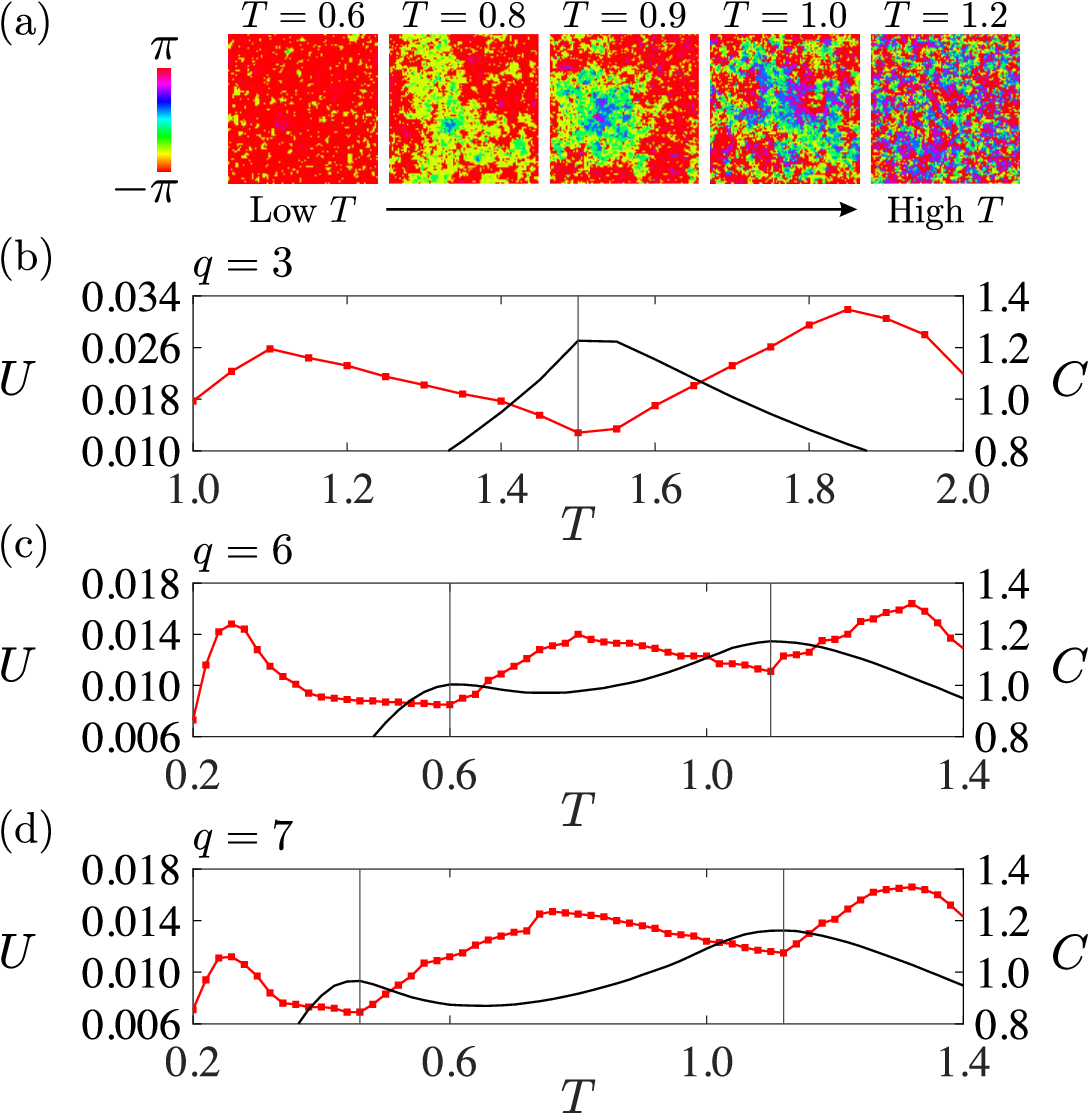}
\par\end{centering}
\caption{\label{fig:Clock}Learning the $q$-state clock model from regression
uncertainty. (a) Typical real-space configurations of the six-state
clock model for instance. (b) Temperature dependence of regression
uncertainty $U(T)$ for reconstructing temperatures $T\in[1,2]$ ($\Delta T=5\times10^{-2}$)
of the three-state clock model with $L=100$. The curve of regression
uncertainty $U(T)$ (red curve) also assumes an M-shape and the valley
appears at $T=1.500\pm0.025$ which matches the critical temperature
$T_{C}=3\slash(2\ln(1+\sqrt{3}))\approx1.492$, suggesting that it
also contains non-trivial information, i.e., the position of the valley,
which can serve as a generic tool for identifying continuous phase
transitions. The black curve corresponds to the temperature dependence
of the system's heat capacity $C(T)$. (c) Temperature dependence
of regression uncertainty $U(T)$ for reconstructing temperatures
$T\in[0.2,1.4]$ ($\Delta T=2\times10^{-2}$) of the six-state clock
model with $L=100$. The curve of regression uncertainty $U(T)$ (red
curve) assumes two successive M-shapes in this case, indicating that
the six-state clock model assumes three different phases, i.e., an
intermediate phase that is absent in its two-state (Ising) and three-state
counterparts. The positions of the two minima of regression uncertainty
$U(T)$ correspond to the two maxima of the system's response function
with respect to the temperature, i.e., the heat capacity $C(T)$ (black
curve), hence not to the BKT transition points in this system. (d)
Analogs of (c), but for the seven-state clock model. See text for
more details.}
\end{figure}

To give a reconstructed temperature $T_{R}$ to these $n_{1}+n_{2}$
indistinguishable samples, one can image that in practice, the NN
could adopt some oversimplified and crude ways if it is totally confused,
e.g., associating all with $T_{1}$ (or $T_{2}$) without distinction
or just randomly guessing. Beyond these bad candidates, a simple but
still decent way for the NN to accomplish its regression task is to
associate the $n_{1}+n_{2}$ indistinguishable samples with their
weighted average temperature $T_{R}=(n_{1}T_{1}+n_{2}T_{2})\slash(n_{1}+n_{2})$.
Indeed, the NN can learn better ways by itself, but if its actual
output $T_{R}$ largely departs from this weighted average, one can
hardly expect an overall good performance of ISP. Further noticing
that in the data generation according to the probability distribution
$\exp(-H\slash T)\slash Z$, stronger thermal fluctuations make those
samples with their energy deviating from the average energy of the
set more easily to be accessed, one thus expects $n_{2}>n_{1}$ since
thermal fluctuations at $T_{2}$ are stronger than $T_{1}$ for $T_{1}<T_{2}<T_{C}$.
Then the weighted average is closer to $T_{2}$ rather than $T_{1}$,
so is the reconstructed temperature $T_{R}$. This thus gives rise
to a smaller regression uncertainty at $T_{2}$ compared to the one
at $T_{1}$, i.e., $U(T_{2})<U(T_{1})$. Similarly, one expects the
opposite for the case $T_{C}<T_{1}<T_{2}$, i.e., $U(T_{1})<U(T_{2})$.
Moreover, the strength of thermal fluctuations at different temperature
$T$ can be characterized by the response property of the system with
respect to the temperature, i.e., by the heat capacity $C(T)=(\langle H^{2}\rangle-\langle H\rangle^{2})\slash(NT^{2})$.
This indicates that a larger heat capacity corresponds to a smaller
regression uncertainty, and in particular, the temperature at which
the heat capacity $C(T)$ reaches its maximum should match exactly
the temperature at which the regression uncertainty $U(T)$ reaches
its minimum. As shown by the heat capacity curve in Fig.~\ref{fig:Ising}(b),
the temperature of the peak of $C(T)$ indeed assumes the same value
as the one of the valley of $U(T)$. Actually, one can also see in
the following that the valley position of regression uncertainty $U(T)$
generally show a consistency with the peak position of heat capacity
$C(T)$, not only in the case of the ferromagnetic Ising model, but
in the cases of three-, six-, and seven-state clock models as well,
supporting the above understanding on the origin of the hidden information
in regression uncertainty.

It is worth mentioning that although the connection between $U(T)$
and $C(T)$ discussed above generally hold true, in practical calculations,
since the temperature region where the regression is performed is
a finite interval, the behavior of $U(T)$ can also be influenced
by the boundary effect of the region. This boundary effect in fact
gives rises to the maxima of $U(T)$ shown in Fig.~\ref{fig:Ising}(b).
Consider for instance the regime below $T_{C}$, one observes that
as the temperature increases, the regression uncertainty $U(T)$ increases,
and reaches a local maximum. The origin of this behavior can be understood
by comparing the regression uncertainty at the left boundary of the
temperature region ($T=2.0$) with the one at the temperature that
is a bit away from the left boundary (e.g., $T=2.02$). $U(T=2.02)$
is larger than $U(T=2.0)$ in this case, since when the NN tries to
reconstruct the temperature at $T=2.02$, it is confused by both the
similar configurations that appear at $T=2.0$ and the ones that appear
at temperatures higher than $T=2.02$, while for reconstructing the
temperature at the left boundary $T=2.0$, it is only confused by
the similar configurations that appear at temperatures higher than
$T=2.0$. As shown in more detailed investigations on this boundary
effect presented in Appendix \ref{sec:Loss-and-maxima} {[}see in
particular Fig.~\ref{fig:Loss-and-maxima}(b){]}, the maxima of $U(T)$
depend on the choice of the temperature region where the regression
is performed, while in sharp contrast, the temperature that corresponds
to the non-trivial minimum of $U(T)$ remains invariant at the critical
temperature $T_{C}$.

\subsection{The generic application of regression uncertainty in learning continuous
phase transitions}

The above investigations on ISP in the ferromagnetic Ising model reveal
an intrinsic connection between regression uncertainty and response
properties of the system. By utilizing this connection, a new type
of unsupervised machine learning approach LFRU is established in this
work for revealing possible phase transitions. To further demonstrate
LFRU as a generic tool for identifying continuous phase transitions,
let us consider the $q=3$ case of the $q$-state clock model $H_{q}=-J\sum_{\langle i,j\rangle}\cos(\theta_{i}-\theta_{j})$,
where $\theta_{i}=2\pi n_{i}\slash q$ denotes the $q$-valued spin
located at the lattice site $i$ with $n_{i}=0,1,\ldots,q-1$. The
generic ability of deep NNs in image recognition enables LFRU to be
directly applied to this model with no additional design. Actually,
the training and testing processes are exactly the same in practice
as the ones involved in studying the ferromagnetic Ising model. Here,
the NN is trained to perform the ISP regression for reconstructing
temperatures $T\in[1,2]$ ($\Delta T=5\times10^{-2}$) of the three-state
clock model, then test its performance and calculate the corresponding
regression uncertainty. As one can see from Fig.~\ref{fig:Clock}(b),
the temperature dependence of regression uncertainty $U(T)$ in this
case also assumes an M-shape. In particular, the valley position $T=1.500\pm0.025$
matches the critical temperature $T_{C}=3\slash(2\ln(1+\sqrt{3}))\approx1.492$
\citep{Savit_RMP_1980,Wu_RMP_1982,Ortiz_NPB_2012} of the continuous
phase transition point of this system.

Remarkably, although the ISP regression algorithm itself is supervised
since there are labels attached to each sample, LFRU can still be
regarded as an unsupervised approach since those labels (e.g., temperature
$T$, rather than phase A/B) do not provide the NN with any prior
physical knowledge about the actual target of LFRU, i.e., possible
phase transitions. As already shown concretely in the ferromagnetic
Ising model and the three-state clock model, one directly trains the
NN to perform regression tasks of the corresponding ISP and then calculates
the regression uncertainty of the well-trained NN afterward {[}see
Eq.~(\ref{eq:Output_variance}) for instance{]}. By the non-trivial
minimum of regression uncertainty, the key information concerning
possible phase transitions is unsupervisedly revealed (see Appendix
\ref{sec:Blank-and-FC} for a blank comparison with no phase transition
involved). For continuous phase transitions, the corresponding response
functions diverge (reach the maximum for finite-size systems) at the
critical point, so the position of the non-trivial minimum of regression
uncertainty directly corresponds to the critical point of phase transitions
{[}see Figs.~\ref{fig:Ising}(b) and \ref{fig:Clock}(b) for instance{]}.
In particular, when applying LFRU, one does not need to calculate
the response functions such as the heat capacity, hence LFRU can assume
practical advantages over conventional approaches in revealing possible
phase transitions of the systems where directly calculating the heat
capacity or other relevant observables is highly non-trivial or even
impossible, e.g., fermion or spin systems with sign problems \citep{Khatami_PRX_2017,Broecker_Sci_Rep_2017,Loh_PRB_1990_SignProblem}.
Moreover, noticing that LFRU does not assume the number of existing
phases in the system under consideration, one could directly use it
to investigate complex physical systems with possible intermediate
phases \citep{Loerting_RMP_2016_water,Digregorio_PRL_2018,Ciamarra_PRL_2020,Zhu_Zheng_PRB_2021,Tianxing_Ma_PRB_2020,Fontenele_PRL_2019}.
As one shall see in the concrete example presented in the following,
LFRU can indeed reveal the key information concerning intermediate
phases.

\section{Reveal intermediate phases from regression uncertainty\label{sec:3phases}}

Let us now discuss how phase transitions in systems with possible
intermediate phases can be investigated by LFRU. To this end, here
the $q=6,7$ cases of the $q$-state clock model are focused for instance.
These models exhibit an intermediate vortex-antivortex condensed BKT
phase between paramagnetic and ferromagnetic phases \citep{Kosterlitz_Rep_Prog_Phys_2016,Kosterlitz_JPhysC_1973,Kosterlitz_JPhysC_1974},
hence in recent years are typically employed to examine the ability
of new machine learning approaches \citep{Miyajima_PRB_2021,Lee_PRE_2019,Scheurer_Nat_Phys_2019,Scheurer_PRL_2020,Jiang_PRB_2020,Wei_PRE_2019}.

To investigate the temperature-driven phase transitions in the six-state
and seven-state clock models, we first train the NN to perform the
ISP regression for reconstructing the temperature $T$ in these two
models, and then calculate their respective regression uncertainty.
As one can see from the temperature dependence of regression uncertainty
$U(T)$ shown in Fig.~\ref{fig:Clock}(c) and Fig.~\ref{fig:Clock}(d),
both the $U(T)$ curves assume two successive M-shapes. Such a structure
with two non-trivial minima, instead of one, is in sharp contrast
to the ones presented in Fig.~\ref{fig:Ising} and Fig.~\ref{fig:Clock}(b),
indicating that the six-state and seven-state clock models assume
three different phases, i.e., an intermediate phase that is absent
in their two-state (Ising) and three-state counterparts. This thus
provides a new type of data-driven evidence on the existence of the
intermediate phase in these two models, corroborating the results
in previous investigations employing conventional approaches such
as the ones utilizing the helicity modulus \citep{Kumano_PRB_2013,Baek_PRE_2013},
the Fisher zeros \citep{Hwang_PRE_2009,Hong_PRE_2020}, the entanglement
entropy of the fixed-point matrix-product-state \citep{Ziqian_Li_PRE_2020},
etc.

At this stage one may incline to associate the two minima of regression
uncertainty $U(T)$ shown in Fig.~\ref{fig:Clock}(c) or Fig.~\ref{fig:Clock}(d)
with the two transition points of the six-state or seven-state clock
model. But thanks to the physical interpretability of LFRU's outputs,
it is clear that the two minima of regression uncertainty $U(T)$
should correspond to the two maxima of the system's response function
with respect to $T$, i.e., the heat capacity $C(T)$, which is indeed
the case as shown by the vertical lines in Figs.~\ref{fig:Clock}(c)
and \ref{fig:Clock}(d). Noticing that the phase transitions associated
with the intermediate phase in these two models are of the BKT type,
i.e., driven by topological defects \citep{Ziqian_Li_PRE_2020,Hostetler_PRD_2021},
the positions of the two maxima of heat capacity $C(T)$ in fact do
not exactly match the critical temperatures \citep{Ziqian_Li_PRE_2020,Hostetler_PRD_2021},
indicating that the positions of the two minima of regression uncertainty
$U(T)$ in Figs.~\ref{fig:Clock}(c) and \ref{fig:Clock}(d) should
not be interpreted as the BKT transition points. This on the one hand
clarifies that quantitatively identifying critical points via LFRU
is restricted to continuous phase transitions, and on the other hand
provides data-driven evidence that the transitions in these systems
are not of the Landau type, corroborating recent investigations \citep{Hong_PRE_2020,Ziqian_Li_PRE_2020,Hostetler_PRD_2021,Miyajima_PRB_2021}.

Finally, another remarkable aspect of LFRU's implementation to highlight
is the robustness against different choices of the NN architecture.
All the machine learning results presented in Fig.~\ref{fig:Ising}
and Fig.~\ref{fig:Clock} are obtained by employing a widely-used
NN that is known as ResNet \citep{Kaiming_He_IEEE_CVPR_2016}. In
Appendix \ref{sec:Training}, the same analysis is performed employing
another type of widely-used NN that is known as GoogLeNet \citep{Szegedy_IEEE_CVPR_2015}.
Their corresponding results agree very well with each other, suggesting
that state-of-the-art developments in the field of artificial intelligence
and data science can be readily exploited to automatically detect
phases of matter in a physically interpretable manner via LFRU.

\section{Conclusions\label{sec:Conclusions}}

The ISP usually has a primary focus on accurately and precisely inferring
system parameters, and thus the uncertainty of regression results
was naturally considered as unfavorable for performing such regression
tasks. However, this unfavorable regression uncertainty actually assumes
an intrinsic connection to the system's response properties, making
it accommodate crucial information concerning possible phase transitions
of the system under consideration. This enables the development of
a fundamentally new type of generic unsupervised machine learning
approach LFRU for automated detection of phases of matter, which is
based on regression instead of classification and produces physically
interpretable results, as manifested concretely in the cases of Ising
model and the three-, six-, and seven-state clock models. The working
mechanism of this approach is quite general, so its application is
not restricted to temperature-driven transitions investigated here.
Moreover, noticing that NNs can easily deal with not only the data
generated from numerical simulations but also experimental data \citep{Rem_Nat_Phys_2019},
including both the real optical imaging \citep{Mulimani_PRR_2020}
and preprocessed physical configurations such as snapshots of the
many-body density matrix \citep{Bohrdt_Nat_Phys_2019}, hence LFRU
is expected to be available for analyzing actual experiments as well.
In addition, various possible phase transitions in a wide range of
nonequilibrium complex many-body systems \citep{Volpe_Nat_Mach_Intell_2020,Chate_Annu_Rev_2020,Munoz_RMP_2018,Volpe_RMP_2016,Vicsek_Phys_Rep_2012},
such as the active matter systems described by the Vicsek model \citep{Vicsek_PRL_1995}
consisting of self-propelled particles, can likewise be investigated
via this approach \citep{Footnote}. In these regards, we believe
that our findings will stimulate further efforts in both developing
and applying physically interpretable machine learning approaches
to unveil new physics in both equilibrium and nonequilibrium complex
many-body systems.
\begin{acknowledgments}
This work was supported by National Science Foundation of China (Grants
No.~11874017 and No.~12275089), Major Basic Research Project of
Guangdong Province (Grant No.~2017KZDXM024), Science and Technology
Program of Guangzhou (Grant No.~2019050001), and the START grant
of South China Normal University.
\end{acknowledgments}

\appendix

\section{Data generation and NN training \label{sec:Training}}

In this work, the data generation for the ferromagnetic Ising model
is done by Monte Carlo simulations using the Swendsen-Wang algorithm
\citep{Swendsen_Wang_PRL_1987,Swendsen_Wang_PhysA_1990} on the Ising
Hamiltonian $H$ with periodic boundary condition imposed. For the
ferromagnetic Ising model with each system size $L=120,140,160,180,200$,
there are $7\times10^{3}$ samples of the steady state spin configuration
generated at each temperature $T$ in the temperature region $[2.0,2.5]$
with the temperature spacing kept as $\Delta T=2\times10^{-2}$, and
in the temperature region $[2.25,2.30]$ with $\Delta T=2\times10^{-3}$,
respectively. These samples are converted to images as shown in Fig.~\ref{fig:Ising}(a),
where the blue or red at each pixel represents the spin $s_{i}$ on
the $i$th lattice site, and then divided into three categories in
the ratio of 5:1:1 forming the training, validation, and test datasets
\citep{Goodfellow_Book_2016}. The data generation for the $q$-state
clock model is done by Monte Carlo simulations on the Hamiltonian
$H_{q}$ with periodic boundary condition imposed. For $q=3$ with
$L=100$, there are $7\times10^{3}$ samples of the steady state spin
configuration generated at each temperature $T$ in the temperature
region $[1,2]$ with $\Delta T=5\times10^{-2}$. For $q=6,7$ with
$L=100$, there are $7\times10^{3}$ samples of the steady state spin
configuration generated at each temperature $T$ in the temperature
region $[0.2,1.4]$ with $\Delta T=2\times10^{-2}$, respectively.
These samples are also converted to images as shown in Fig.~\ref{fig:Clock}(a),
where the cyclic color at each pixel represents the planar angle $\theta_{i}$
on the $i$th lattice site, and then divided into three categories
in the same ratio as above forming the datasets.

\begin{figure}
\noindent \begin{centering}
\includegraphics[width=3.3in]{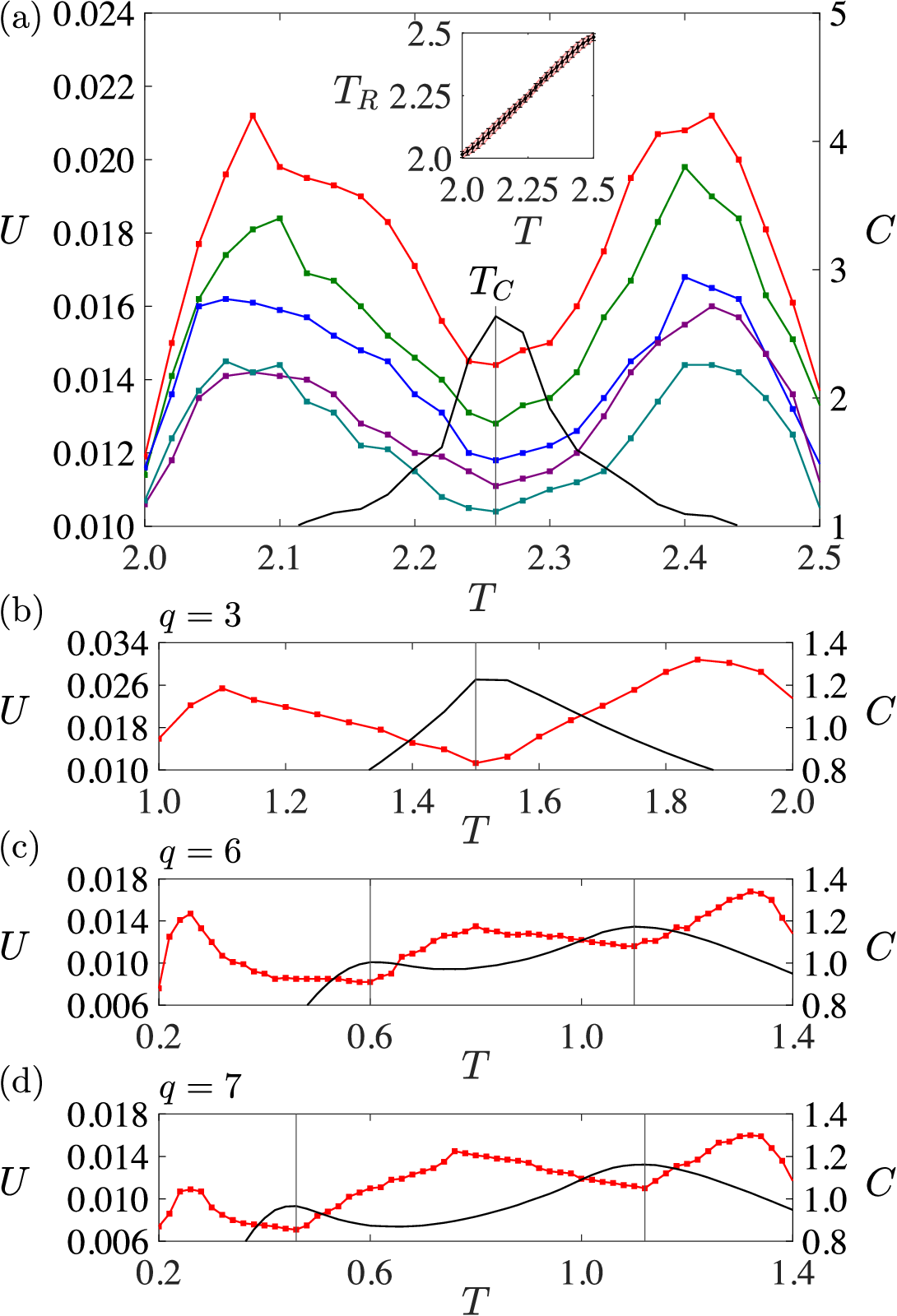}
\par\end{centering}
\caption{\label{fig:ggn}Analogs of Fig.~\ref{fig:Ising}(b) and Figs.~\ref{fig:Clock}(b-d),
but obtained by employing GoogLeNet instead of ResNet. These two employed
NNs have very different designs for the feature extraction, but their
results match very well, manifesting that the implementation of LFRU
is robust against different choices of the NN architecture. See text
for more details.}
\end{figure}

Using these datasets, two widely-used NNs are employed, namely, ResNet
\citep{Kaiming_He_IEEE_CVPR_2016} and GoogLeNet \citep{Szegedy_IEEE_CVPR_2015},
to perform regression tasks of ISP. The results obtained by ResNet
are shown in Fig.~\ref{fig:Ising} and Fig.~\ref{fig:Clock} in
the main text, and the results obtained by GoogLeNet are shown in
Fig.~\ref{fig:ggn}. These NNs are similarly trained by using the
Adam optimizer \citep{Kingma_ICLR_2015} traversing the training dataset
for 10 epochs at a learning rate $\alpha=10^{-3}$, together with
a validation at each epoch traversing the validation dataset. After
that, the samples in the test datasets are used to calculate the regression
uncertainty $U(T)$. All the machine learning results in this work
are averaged over 20 independent training and testing processes.

\section{Loss function, and the maxima of regression uncertainty\label{sec:Loss-and-maxima}}

\begin{figure}
\noindent \begin{centering}
\includegraphics[width=3.3in]{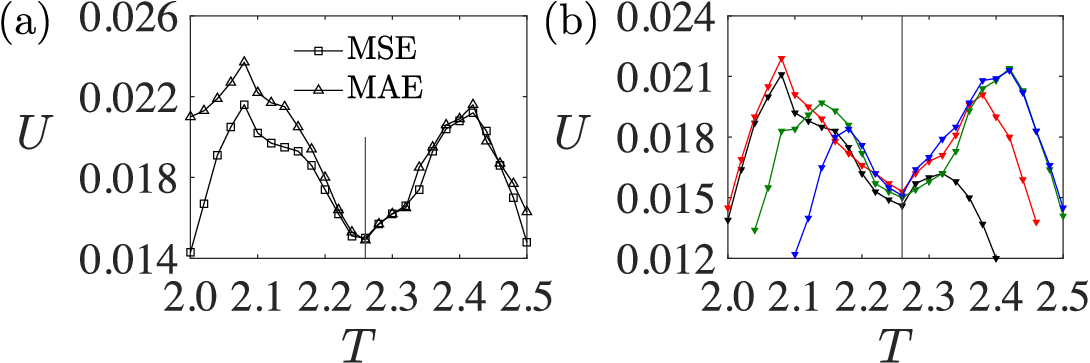}
\par\end{centering}
\caption{\label{fig:Loss-and-maxima}(a) Temperature dependence of regression
uncertainty $U(T)$ for reconstructing temperatures $T\in[2.0,2.5]$
($\Delta T=2\times10^{-2}$) of Ising model with $L=120$ obtained
by using the mean square error (MSE) and the mean absolute error (MAE)
loss functions to train ResNet, respectively. Their valley positions
of regression uncertainty $U(T)$ are numerically the same, suggesting
that LFRU's implementation is robust against different reasonable
choices of the loss function. (b) Temperature dependence of regression
uncertainty $U(T)$ for reconstructing temperatures $T\in[2.00,2.40]$
(corresponding to the black curve), $T\in[2.00,2.46]$ (red), $T\in[2.04,2.50]$
(green), and $T\in[2.10,2.50]$ (blue), respectively, of Ising model
with $L=120$, $\Delta T=2\times10^{-2}$. Maxima of regression uncertainty
$U(T)$ depend on the choice of the temperature region where the regression
is performed, while in sharp contrast, the temperature that corresponds
to the non-trivial minimum of $U(T)$ remains invariant at $T=2.26\pm0.01$.
These results suggest that the boundary effect of the parameter region
only imposes noticeable influences on the maxima positions of regression
uncertainty, while as long as there exists at least two phases in
the parameter region where the regression is performed, the LFRU's
implementation and the valley positions of regression uncertainty,
i.e., the critical point predicted via LFRU, are robust against different
reasonable choices of the parameter region. See text for more details.}
\end{figure}

In the main text, the mean square error (MSE) loss function $\mathbb{L}=\langle(T-T_{R})^{2}\rangle$
is employed to train the NN for instance. Here, for the sake of completeness,
the same analysis is performed employing the mean absolute error (MAE)
loss function $\mathbb{L}=\langle\vert T-T_{R}\vert\rangle$, which
is another widely-used loss function that is suitable for regression
tasks. Their results are presented in Fig.~\ref{fig:Loss-and-maxima}(a).
For reconstructing temperatures $T\in[2.0,2.5]$ ($\Delta T=2\times10^{-2}$)
of Ising model with $L=120$, these two different choices of loss
function give rise to similar behavior of regression uncertainty.
In particular, the corresponding valley positions of regression uncertainty
$U(T)$ are numerically the same. This thus indicates that the signals
generated by LFRU concerning possible phase transitions are robust
against different choices of the loss function, as long as the NN
trained with the loss function can achieve a good performance in reconstructing
the system parameter.

Moreover, to better understand the origin of the maxima of regression
uncertainty, the boundary effect of the parameter region where the
regression is performed is investigated. It can be seen clearly from
Fig.~\ref{fig:Loss-and-maxima}(b) by comparing the behaviors of
regression uncertainty $U(T)$ for respectively reconstructing temperatures
$T\in[2.00,2.40]$, $T\in[2.00,2.46]$, $T\in[2.04,2.50]$, $T\in[2.10,2.50]$
of Ising model. Although the valley of $U(T)$ in these cases remains
invariant at the critical temperature $T_{C}$, the hills of $U(T)$
are indeed sensitive to such slight changes of the temperature region,
hence their corresponding temperatures do not assume a particular
physical significance.

\section{Finite-size analysis\label{sec:Finite_size_analysis}}

\begin{figure}
\noindent \begin{centering}
\includegraphics[width=3.3in]{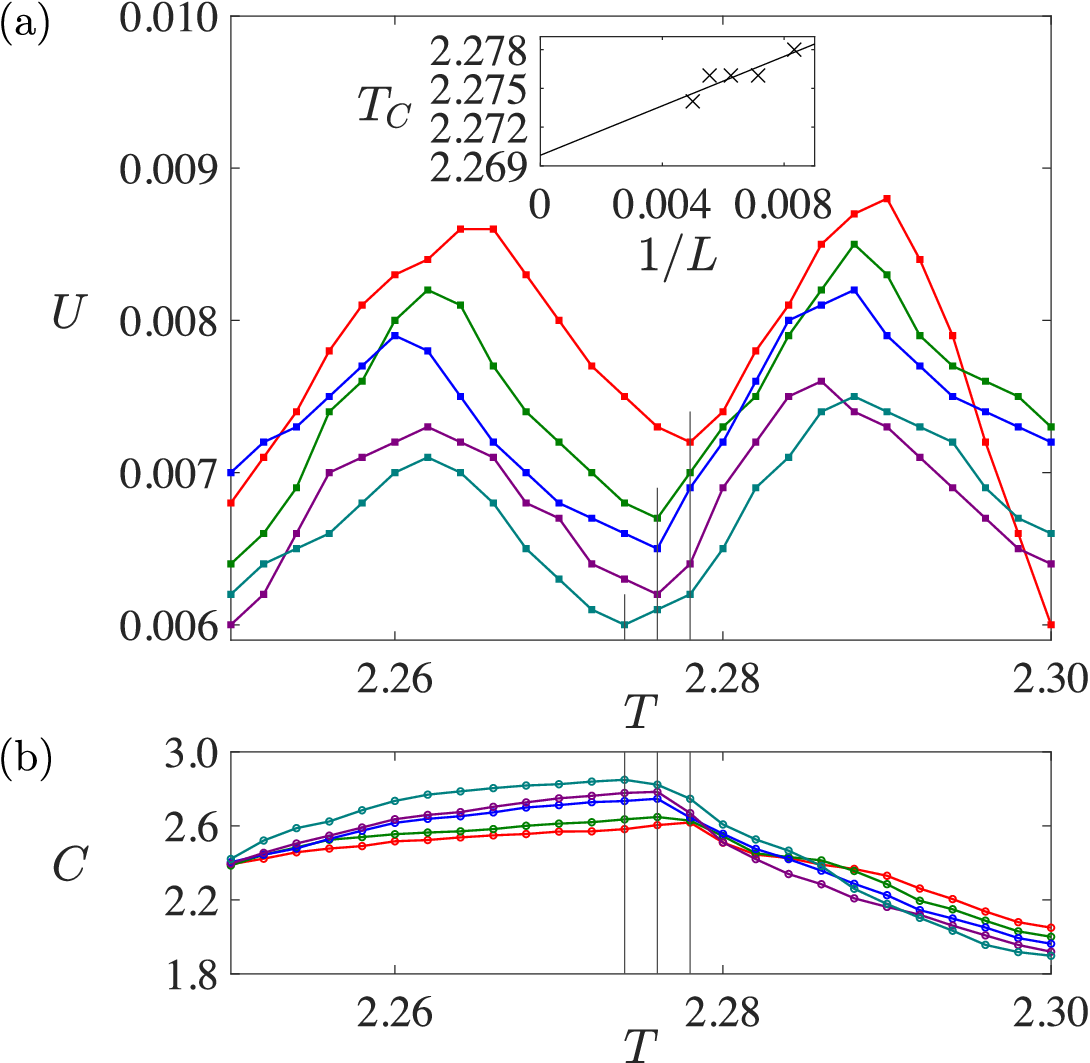}
\par\end{centering}
\caption{\label{fig:Finite_size_effect}(a) Temperature dependence of regression
uncertainty $U(T)$ for reconstructing temperatures $T\in[2.25,2.30]$
($\Delta T=2\times10^{-3}$) of Ising model with $L=120$ (red curve),
$L=140$ (green), $L=160$ (blue), $L=180$ (purple), and $L=200$
(cyan). The valley of $U(T)$ appears at $T=2.278\pm0.001$ for $L=120$,
at $T=2.276\pm0.001$ for $L=140,160,180$, at $T=2.274\pm0.001$
for $L=200$. Inset: Fitting of valley position $T_{C}(L)$ of regression
uncertainty $U(T)$ as a function of $1\slash L$, which leads to
an estimation of $T_{C}=2.270\pm0.001$ for the critical temperature
in the thermodynamic limit. (b) System's heat capacity $C(T)$ within
$[2.25,2.30]$ ($\Delta T=2\times10^{-3}$) with different system
sizes. These results indicate that LFRU is applicable for both the
extensive search for phase transitions within a relatively large parameter
ranges (corresponding to relatively large $\Delta T$), as well as
for the quantitative analysis within the vicinity of critical points
(corresponding to relatively small $\Delta T$). See text for more
details.}
\end{figure}

Temperatures $T\in[2.0,2.5]$ ($\Delta T=2\times10^{-2}$) of Ising
model are reconstructed in Fig.~\ref{fig:Ising}(b) in the main text.
Here, a detailed check in the vicinity of $T_{C}$ with a higher resolution
is performed. As one can see from Fig.~\ref{fig:Finite_size_effect}(a),
for reconstructing temperatures $T\in[2.25,2.30]$ with $\Delta T=2\times10^{-3}$,
the temperature dependence of regression uncertainty $U(T)$ also
assumes the M-shape. According to its valley, the predicted critical
temperature $T_{C}(L)$ for each system size is $T_{C}(L=120)=2.278\pm0.001$,
$T_{C}(L=140,160,180)=2.276\pm0.001$, $T_{C}(L=200)=2.274\pm0.001$.
The maximum position of $C(T)$ shown in Fig.~\ref{fig:Finite_size_effect}(b)
shifts among different system sizes due to finite-size effects, but
the intrinsic connection between regression uncertainty and response
properties ensures that the position of the valley of $U(T)$ in Fig.~\ref{fig:Finite_size_effect}(a)
keeps matching the maximum position of $C(T)$ irrespective of the
system size. These finite-size data of $T_{C}(L)$ are fitted as a
function of $1\slash L$ as shown in the inset of Fig.~\ref{fig:Finite_size_effect}(a),
which leads to an estimation of $T_{C}=2.270\pm0.001$ (the resolution
is determined by $\Delta T$) for the critical temperature in the
thermodynamic limit.

\begin{figure}
\noindent \begin{centering}
\includegraphics[width=3.3in]{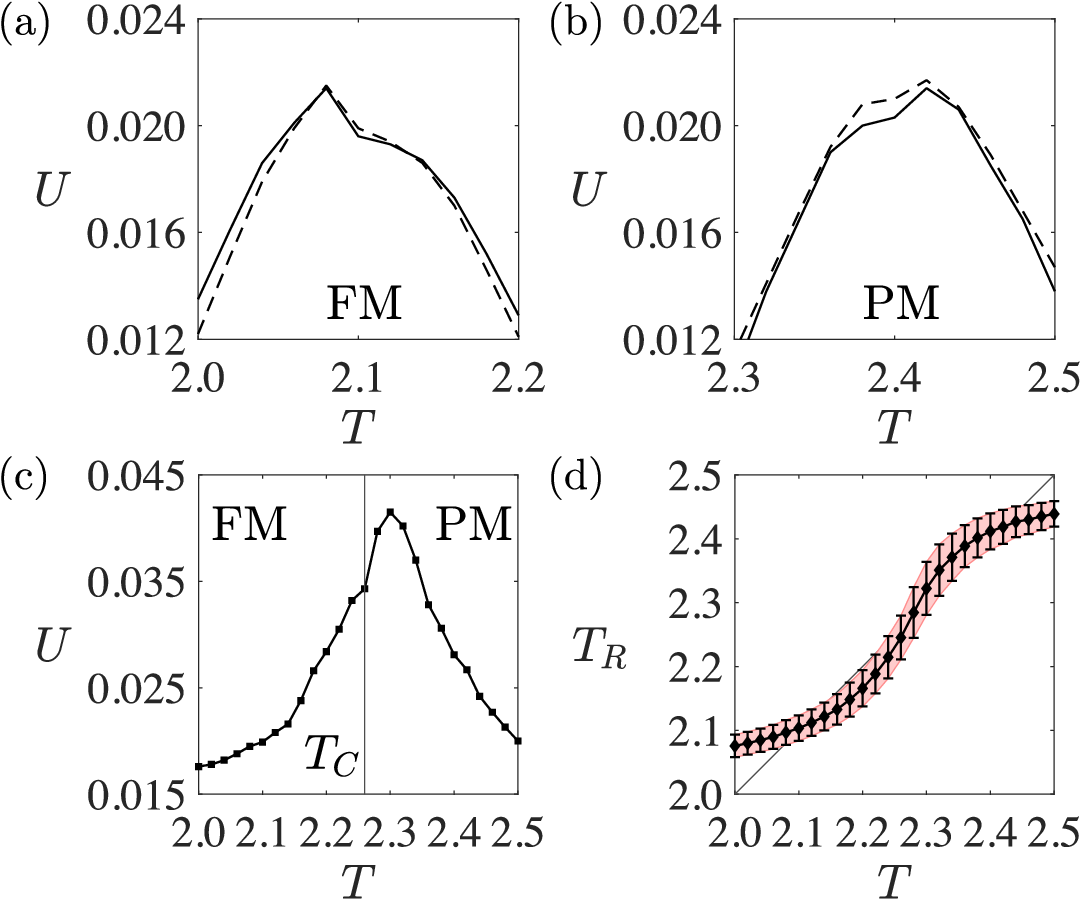}
\par\end{centering}
\caption{\label{fig:Blank-and-FC}(a) Temperature dependence of regression
uncertainty $U(T)$ for reconstructing temperatures $T\in[2.0,2.2]$
($\Delta T=2\times10^{-2}$) of Ising model with $L=120$. Only the
ferromagnetic (FM) phase exists in this temperature region, so the
curve of $U(T)$ assumes no non-trivial minimum. The solid curve and
the dashed curve represent the results obtained by ResNet and GoogLeNet,
respectively. (b) Analogs of (a), but concerning the high-temperature
region $[2.3,2.5]$ where only the paramagnetic (PM) phase exists.
These results confirm that LFRU does not require prior system-specific
knowledge about the number of existing phases as a given condition.
(c) Temperature dependence of regression uncertainty $U(T)$ for reconstructing
temperatures $T\in[2.0,2.5]$ ($\Delta T=2\times10^{-2}$) of Ising
model with $L=120$ obtained by using a very simple fully-connected
NN that assumes a deliberately produced poor performance on ISP. This
curve of $U(T)$ looks similar to the ones presented in (a) and (b)
with no phase transition involved, suggesting that the NN's good performance
on ISP is a prerequisite of LFRU for automated detection of phases
of matter. (d) Regression results corresponding to (c). The reconstructed
temperature $T_{R}$ largely departs from the target $T$ (the diagonal
line represents the ideal regression results $T_{R}=T$ for the ISP).
The error bars and the range of the colored error band represent the
regression uncertainty $U(T)$, but here the regression uncertainty
no longer contain hidden information that can be utilized to reveal
possible phase transitions, since this NN fails to accomplish its
regression task. See text for more details.}
\end{figure}

\section{Blank comparison and the performance on ISP\label{sec:Blank-and-FC}}

\begin{figure}
\noindent \begin{centering}
\includegraphics[width=3.3in]{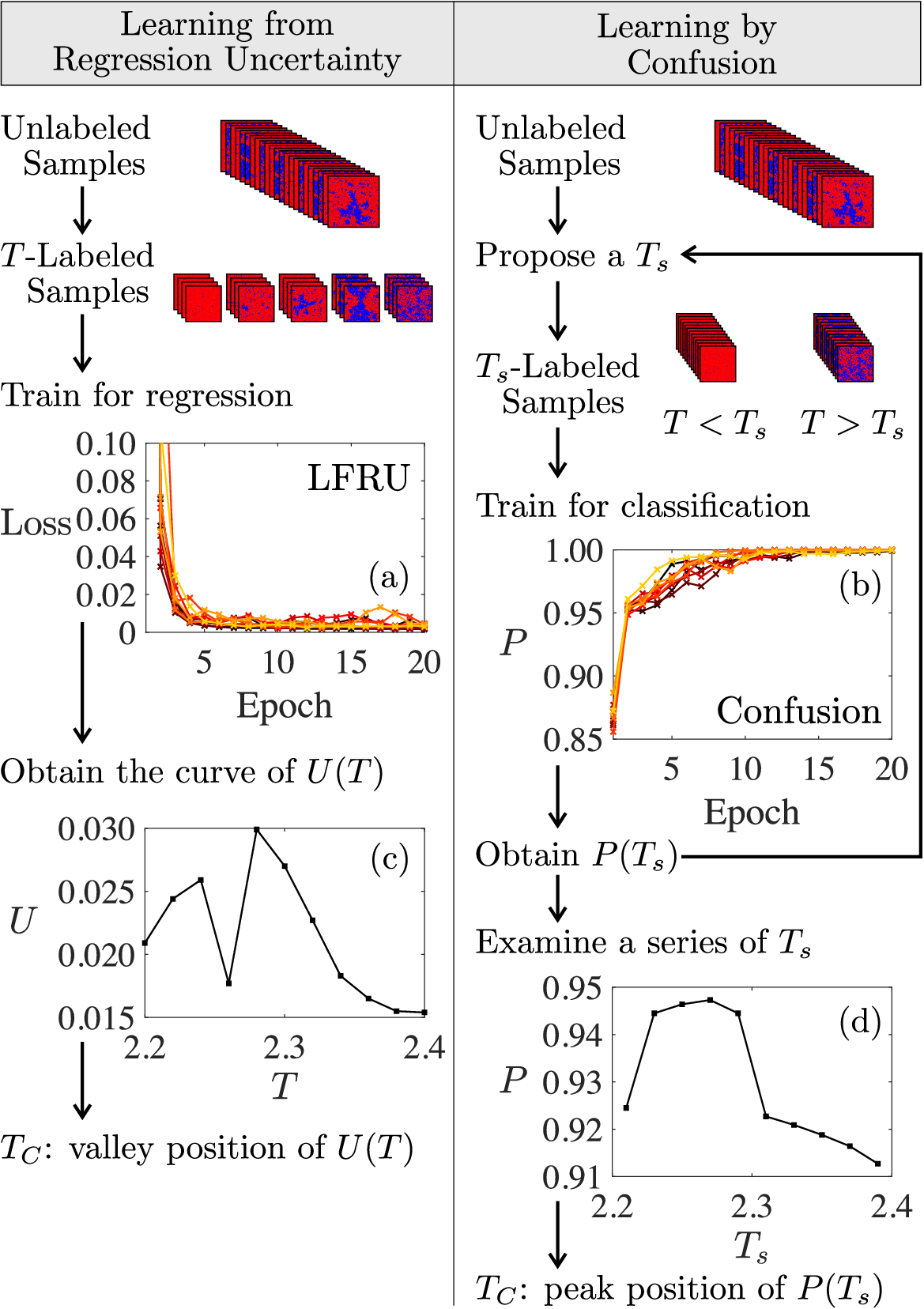}
\par\end{centering}
\caption{\label{fig:Confusion}Comparison of LFRU and the learning-by-confusion
approach. Left panel: Illustration of applying LFRU for identifying
$T_{C}$ of the ferromagnetic Ising model $(L=120)$ with samples
generated at $T\in[2.2,2.4]$ with $\Delta T=2\times10^{-2}$. Right
panel: Illustration of applying the learning-by-confusion approach
for identifying the same $T_{C}$ with the same datasets. (a) The
training process of the NN for reconstructing temperatures, traversing
the training dataset for $20$ epochs. Different colors represent
different independent trajectories of training. (b) The training process
of the NN for classifying phase A ($T<T_{s}$) and phase B ($T>T_{s}$),
traversing the same dataset for $20$ epochs. These results show that
comparing the implementation of LFRU and the learning-by-confusion
approach, they require similar numbers of epochs to give birth to
a well-trained NN. (c) Temperature dependence of regression uncertainty
$U(T)$ obtained from the well-trained NNs in (a), averaged over $10$
independent trajectories. (d) $T_{s}$ dependence of the classification
accuracy $P(T_{s})$ obtained from the well-trained NNs in (b), averaged
over $10$ independent trajectories. See text for more details.}
\end{figure}

For a blank comparison, LFRU is applied here on the ferromagnetic
Ising model concerning solely the low-temperature region $[2.0,2.2]$
($\Delta T=2\times10^{-2}$) employing both ResNet \citep{Kaiming_He_IEEE_CVPR_2016}
and GoogLeNet \citep{Szegedy_IEEE_CVPR_2015}, respectively. As one
can see in Fig.~\ref{fig:Blank-and-FC}(a), the temperature dependence
of regression uncertainty $U(T)$ assumes a single maximum and no
non-trivial minimum, namely, only the boundary effect of the parameter
region exhibits signs. This is consistent with the fact that only
one phase, i.e., the ferromagnetic phase, exists with no phase transition
in the temperature region $[2.0,2.2]$. Similar results are obtained
concerning solely the high-temperature region $[2.3,2.5]$ ($\Delta T=2\times10^{-2}$)
that corresponds to only the paramagnetic phase, as shown in Fig.~\ref{fig:Blank-and-FC}(b).
These results verify that the non-trivial minimum of regression uncertainty
in performing regression tasks of ISP is indeed an unsupervised signal
for revealing possible phase transitions of the system under consideration.

It is also worth mentioning that the signals generated by LFRU concerning
possible phase transitions are based on the NN's good performance
on ISP. Such a performance is not guaranteed by an arbitrary NN. Here,
a poor performance is deliberately produced {[}see~Fig.~\ref{fig:Blank-and-FC}(d){]}
concerning the temperature region $[2.0,2.5]$ ($\Delta T=2\times10^{-2}$)
on the ferromagnetic Ising model with $L=120$. This is obtained by
using a very simple fully-connected NN, which consists of an input
layer with $3L^{2}$ neurons (3 for RGB), a hidden layer with $L$
neurons (rectified linear unit as the activation function, batch normalization
applied) \citep{Goodfellow_Book_2016}, and an output layer with only
one neuron. It is worth mentioning that the fully-connected NN might
also perform well after fine-tuning, but a poor performance is needed
here for technical discussions. As one can see from Fig.~\ref{fig:Blank-and-FC}(c),
when the NN cannot well accomplish its regression task of ISP, the
resulting behavior of regression uncertainty $U(T)$ looks similar
to the ones presented in Figs.~\ref{fig:Blank-and-FC}(a) and \ref{fig:Blank-and-FC}(b)
with no phase transition involved. However, the temperature region
under consideration in Fig.~\ref{fig:Blank-and-FC}(c) is indeed
across $T_{C}$, suggesting that one cannot distinguish whether a
single peak of $U(T)$ is associated with a phase transition or not
merely by monitoring the regression uncertainty $U(T)$ itself. A
reasonable judgment is that any investigation according to the information
extracted from a poor performance will suffer from a lack of legitimacy.
Therefore, the NN's good performance on ISP is a prerequisite of LFRU
for automated detection of phases of matter.

\section{Comparison with the learning-by-confusion approach}

Here in this appendix, LFRU is directly compared with the learning-by-confusion
approach \citep{van_Nieuwenburg_Nat_Phys_2017}, which is also an
unsupervised learning approach for automated detection of phases of
matter. As a concrete example, let us consider identifying $T_{C}$
of the ferromagnetic Ising model with samples generated at temperatures
$T\in[2.2,2.4]$ with the temperature spacing $\Delta T=2\times10^{-2}$.
As illustrated in the right panel of Fig.~\ref{fig:Confusion}, when
applying the learning-by-confusion approach, one shall first propose
a possible critical point $T_{s}$. After that, the samples satisfying
$T<T_{s}$ are accordingly labeled with $(\tilde{y}_{A}=1,\tilde{y}_{B}=0)$
as phase A, and the samples satisfying $T>T_{s}$ are labeled with
$(\tilde{y}_{A}=0,\tilde{y}_{B}=1)$ as phase B. Then the NN is trained
to perform an NN-based binary classification between these artificial
phases, outputting for each sample its confidences $(y_{A},y_{B})$
of identifying the sample as phase A or B. The loss function should
be suitable for the binary classification task, e.g., the cross-entropy
function $\mathbb{L}^{\prime}=\langle-\tilde{y}_{A}\ln y_{A}-\tilde{y}_{B}\ln y_{B}\rangle$
\citep{Goodfellow_Book_2016}, instead of the MSE or MAE loss function
for regression. After training, one obtains the classification accuracy
$P(T_{s})$ that the NN successfully matches the samples in the test
dataset with their attached labels corresponding to the proposed $T_{s}$.
According to Ref.~\citep{van_Nieuwenburg_Nat_Phys_2017}, one shall
examine a series of different $T_{s}$, and the position of the maximum
of $P(T_{s})$ is considered the critical point $T_{C}$ predicted
via the learning-by-confusion approach. Similar to LFRU, the resolution
of $T_{C}$ predicted via the learning-by-confusion approach is also
determined by $\Delta T$. The NN used by the learning-by-confusion
approach and the NN used by LFRU only have slight differences in their
output layers, i.e., there are two neurons in the former for binary
classification, but only one single neuron in the latter for regression,
and hence the time it takes to train these NNs traversing the same
datasets for an epoch is approximately the same. Figs.~\ref{fig:Confusion}(a)
and (b) show that the convergence speed is also similar, so they require
similar numbers of epochs to give birth to a well-trained NN. However,
a well-trained NN for the learning-by-confusion approach can just
give a single $P(T_{s})$, and if there are $m$ different possible
values of $T_{s}$ to be examined as $T_{C}$ (at least $m\geqslant3$,
and usually much larger), it requires $m$ well-trained NNs. In sharp
contrast, when applying LFRU, a complete curve of $U(T)$ is directly
obtained from one well-trained NN, and the valley position of $U(T)$
can be considered as the critical point $T_{C}$ predicted via LFRU.
Therefore, the approach developed in this work could be about $m$
times faster than the widely-used learning-by-confusion approach.
The direct comparison of their results is presented in Figs.~\ref{fig:Confusion}(c)
and (d).

\end{document}